# C Implementation & comparison of companding & silence audio compression techniques

Mrs. Kruti Dangarwala[1] and Mr. Jigar Shah[2]

[1]Department of Computer Engineering,
Sri S'ad Vidya Mandal Institute of Technology
Bharuch, Gujarat, India

[2]Department of Electronics and Telecommunication Engineering
Sri S'ad Vidya Mandal Institute of Technology
Bharuch, Gujarat, India

**Abstract**

Just about all the newest living room audio-video electronics and PC multimedia products being designed today will incorporate some form of compressed digitized-audio processing capability. Audio compression reduces the bit rate required to represent an analog audio signal while maintaining the perceived audio quality. Discarding inaudible data reduces the storage, transmission and compute requirements of handling high-quality audio files. This paper covers wave audio file format & algorithm of silence compression method and companding method to compress and decompress wave audio file. Then it compares the result of these two methods.

## 1. Introduction

Audio compression reduces the bit rate required to represent an analog audio signal while maintaining the perceived audio quality. Most audio decoders being designed today are called "lossy," meaning that they throw away information that cannot be heard by most listeners. The information to be discarded is based on psychoacoustics, which uses a model human auditory perception to determine which parts of the audible spectrum the largest portion of the human population can detect. First, an audio encoder [1] divides the frequency domain of the signal being digitized into many bands and analyzes a block of audio to determine what's called a "masking threshold." The number of bits used to represent a tone depends on the masking threshold. The noise associated with using fewer bits is kept low enough so that it will not be heard. Tones that are completely masked may not have any bits allocated to them. Discarding inaudible data reduces the storage, transmission and compute requirements of handling high-quality audio files. Consider the example of a typical audio signal found in a CD-quality audio device. The CD player produces two channels of audio. Each analog signal [2] in each channel is sampled at a 44.1-kHz sample rate. Each sample is represented as a 16-bit digital data word. To produce both channels requires a data rate of 1.4 Mbits/second. However, with audio compression this data rate is reduced around an order of magnitude. Thus, a typical CD player is reading compressed data from a compact disk at a rate just over 100 Kbits/s. Audio compression really consists of two parts. The first part, called *encoding*, transforms the digital audio data that resides, say, in a WAVE file, into a highly compressed form called bitstream. To play the bitstream on your soundcard, you need the second part, called *decoding*. Decoding takes the bitstream and re-expands it to a WAVE file.

## 2. WAVE AUDIO FILE FORMAT

The WAVE file format [1] is a subset of Microsoft's RIFF spec, which can include lots of different kinds of data. RIFF is a file format for storing many kinds of data, primarily multimedia data like audio and video. It is based on *chunks* and *sub-chunks*. Each chunk has a type, represented by a four-character tag. This chunk type comes first in the file, followed by the size of the chunk, then the contents of the chunk. The entire RIFF file is a big chunk that contains all the other chunks. The first thing in the contents of the RIFF chunk is the "form type," which describes the overall type of the file's contents. So the structure of wave audio file looks like this:   a*)* RIFF Chunk b*)* Format Chunk c*)* Data Chunk

Table 1:　RIFF CHUNK

| Byte Number | Description |
|---|---|
| 0-3 | "RIFF"(ASCII Character) |
| 4-7 | Total Length of Package to follow (Binary) |
| 8-11 | "WAVE"(ASCII Character) |

Description of RIFF chunk as follows:
```
Offset   Length              Contents
  0      4 bytes             'RIFF'
  4      4 bytes             <file length - 8>
  8      4 bytes             'WAVE'
```

Table 2: FORMAT CHUNK

| Byte Number | |
|---|---|
| 0-3 | "fmt_"(ASCII character) |
| 4-7 | Length of format chunk(Binary) |
| 8-9 | Always 0x01 |
| 10-11 | Channel nos(0x01=mono,0x02=stereo) |
| 12-15 | Sample Rate(Binary,in Hz) |
| 16-19 | Bytes Per Second |
| 20-21 | Bytes Per Sample 1=8-bit mono,2=8-bit stereo/16-bit mono,4=16-bit stereo |
| 22-23 | Bits Per Sample |

Description of FORMAT chunk as follows:
```
12   4 bytes  'fmt '
16   4 bytes  0x00000010
20   2 bytes  0x0001     // Format tag: 1 = PCM
22   2 bytes  <channels>
24   4 bytes  <sample rate>   // Samples per second
28   4 bytes  <bytes/second>  // sample rate * block
                                 align
32   2 bytes  <block align>   // channels *
             bits/sample / 8    34   2 bytes
             <bits/sample>   // 8 or 16
```

Table 3: DATA CHUNK

| Byte Number | |
|---|---|
| 0-3 | "data"(ASCII character) |
| 4-7 | Length of data to follow |
| 8-end | Data(Samples) |

Description of DATA chunk as follows:
```
36    4 bytes 'data'
40    4 bytes <length of the data block>
44    bytes <sample data>
```

The sample data must end on an even byte boundary. All numeric data fields are in the Intel format of low-high byte ordering. 8-bit samples are stored as unsigned bytes, ranging from 0 to 255. 16-bit samples are stored as 2's-complement signed integers, ranging from -32768 to 32767.

For multi-channel data, samples are interleaved between channels, like this: sample 0 for channel 0, sample 0 for channel 1, sample 1 for channel 0 ,sample 1 for channel 1. For stereo audio, channel 0 is the left channel and channel 1 is the right.

## 3. Silence Compression & Decompression Techniques

### 3.1 Introduction

Silence Compression [4] on sound files is the equivalent of run length encoding on normal data files. In this case, the Runs we encode are sequences of relative silence in a sound file. Here we replace sequences of relative silence with absolute silence. So it is known as Lossy technique.

### 3.2 User Parameters:

1) Thresold Value *:* It considered as Silence. With 8-bit sample 80H considered as "pure" silence. Any Sample value within a range of plus or minus 4 from 80H considered as silence.

2) Silence_Code*:* It is code to encode a run of silence. We used value FF to encode silence.The Silence_code is followed by a single byte that indicates how many consecutive silence codes there are.

3) Start_Threshold*:* It recognize the start of a run of silence. We would not want to start encoding silence after seeing just a single byte of silence. It does not even become economical until 3 bytes of silence are seen. We may want to experiment with even higher values than 3 to see how it affects the fidelity of the recoding.

4) Stop_Threshold*:* It indicates how many consecutive non silence codes need to be seen in the input stream before we declare the silence run to be over.

### 3.3 Silence Compression Algorithm (Encoder)

1) Read 8-bit Sample Data From audio file.
2) Checking of Silence means find atleast 5 consecutive silence value: 80H or +4 /- 4 from 80H.  (Indicate start of silence)
3) If get, Encode with Silence_Code followed by runs. \ (Consecutive Silence values).
4) Stop to Encode when found atleast two Non-Silence values.
5) Repeat all above steps until end of file character found.
6) Print input File size, Output File Size and Compression Ratio.

This algorithm [4] takes 8-bit wave audio file as input. Here It find starting of silence means check that at least 5 consecutive silence value present or not. Here 80H considered as pure silence and +/- 4 from 80H also consider as silence. If found, then it start encoding process. Here consecutive silence values are encoded by silence_code followed by runs (Consecutive silence values). It stop encoding when it found at least two non-silence values. Then it generate compressed file extension of that file is also wav file.
Example of algorithm as follows:

Input file consists of sample data like 80 81 80 81 80 80 80 45. Output file consists of compressed data like FF745.
It display following attributes of input wave audio file.
   a. Input file size in terms of bytes
   b. Output file size in terms of bytes
   c. Compression ratio.

3.4 Silence Dcompression Algorithm (Decoder)

1) Read 8-bit Sample from Compress file.
2) Check the Silence code means 0xff if it found, Check the next value means runs which Indicate no of silence value.
3) Replace it with 0x80 (silence value) no of runs times.
4) Repeat above step until we get end of file Character.

Example of algorithm as follows:

In input file (compressed file (extension of this file .wav)) we find silence code , if we get then we check next value which indicate no of silence value. Then we replace with silence value no of runs times decided by user. We stop the procedure when we get end of file character.

If we get value 0xff5 in compress file, decode that value by
0x80 0x80 0x80 0x80 0x80 0x80

## 4. Companding Compression & Decompression Techniques

4.1 Introduction

Companding [4] uses the fact that the ear requires more precise samples at low amplitudes (soft sounds). But is more forgiving at higher amplitudes. A typical ADC used in sound cards for personal computers convert voltages to numbers linearly. If an amplitude a is converted to the number n, then amplitude 2a will be converted to the number 2n. It examines every sample in the sound file and uses a nonlinear formula to reduce the no. of bits devoted to it.

Non-Linear Formula For 16-bit samples to 15-bit samples conversion:

**Mapped=32767*(pow (2, sample/65356)     (1)**

Using this formula every 16-bit sample data converted into 15-bit sample data. It performs non-linear mapping such that small samples are less affected than large ones.

Mapped 15-bit numbers can be decoded back into the original 16-bit samples by the inverse formula:

**Sample=65356 log2 (1+mapped/32767)     (2)**

Reducing 16-bit numbers to 15-bits does not produce much compression. Better Compression can be achieved by substituting a smaller number for 32767 in "(1)" & "(2)". A value of 127, For example would map each 16-bit sample into 8-bit sample. So in this case we compress file with compression ratio of 0.5 means 50%. Here Decoding should be less accurate. A 16-bit sample of 60,100, for example, would be mapped into the 8-bit number 113, but this number would produce 60,172 when decoded by "(2)". Even worse, the small 16-bit sample 1000 would be mapped into 1.35, which has to be rounded to 1. When "(2)" is used to decode a 1, it produce 742, significantly different from the original sample.

Here the Amount of Compression[2] should thus be a user-Controlled parameter. And this is an interesting example of a compression method where the compression ratio is known in advance. Now no need to go through the "(1)" & "(2)". Since the mapping of all the samples can be prepared in advance in a table. Both decoding and encoding are thus fast.Use [4] in this method.

4.2 Companding Compression Algorithm

1) Input the No. of Bits to use for output code.
2) Make Compress Look-Up Table using Non-Linear Formula : (8-bit to Input bit)
   Value=128.0*(pow (2, code/N)-1.0)+0.5
   Where, code: pow(2,inputbit) to 1
   N: pow (2, Inputbit)
   For each code we assign value 0 to 15 in table:
   Index of table           value
   J+127      →             code+N-1
   128-j      →             N-code
   where j=value to zero
3) Now Read 8-bit samples from audio file and That sample become the index of compress look-up table, Find corresponding value, that output vale store in output file.
4) Repeat step-III until we get end of file character.
5) Print the Input file size in bytes, output file size in bytes & compression ratio.

Description of algorithm is that it used for converting 8- bit sample file into user defined output bit. For example if we input output bit : 4 then we achieve 50% compression. and we say compression ratio is 0.5.
So we say that in this method compression ratio is known in advance. We adjust compression ratio according our requirement. So it is crucial point compare to another method.

### 4.3 Companding Decompression Algorithm

1) Find the No. of bits used in compressed file.
2) Make Expand Look-Up table using Non-Linear Formula:
   (Input bit → 8-bit)
   Value=128.0*(pow (2.0, code/N)-1.0)+0.5
   Where code: 1 to pow (2, Inputbit)
   N: pow (2, Inputbit)

   For each code: we assign value 0 to 255 in table:
   Index of table          value
   N+code-1 →         128+(value+last_value)/2
   N-code →              127-(value+last_value)/2

   Here initially, last_value=0 & for each code, last_value=value

3) Now read input bit samples from audio file & that sample become the index of expand look-up table, find corresponding value, that output sample value store in output file.
4) Repeat the step-III until file size becomes zero.

## 5. Result/comparisons Between Two Lossy Method

### 5.1 Companding Compression Method

1) INPUT AUDIO FILE:

Name of File:      J1. WAV
Media Length:   4.320 sec
Audio Format:   PCM, 8000Hz, 8-Bit, Mono
File Size:             33.8KB (34,618 BYTES)

| User Parameter (No. Of Bits) | 1 |
|---|---|
| Input File Size(in Bytes) | 34618 |
| Output File Size(in Bytes) | 4390 |
| Compression Ratio(in Percentage) | 88% |

| User Parameter (No. Of Bits) | 2 |
|---|---|
| Input File Size(in Bytes) | 34618 |
| Output File Size(in Bytes) | 8709 |
| Compression Ratio(in Percentage) | 75% |

| User Parameter (No. Of Bits) | 3 |
|---|---|
| Input File Size(in Bytes) | 34618 |
| Output File Size(in Bytes) | 13028 |
| Compression Ratio(in Percentage) | 63% |

| User Parameter (No. Of Bits) | 4 |
|---|---|
| Input File Size(in Bytes) | 34618 |
| Output File Size(in Bytes) | 17347 |
| Compression Ratio(in Percentage) | 50% |

| User Parameter (No. Of Bits) | 5 |
|---|---|
| Input File Size(in Bytes) | 34618 |
| Output File Size(in Bytes) | 21666 |
| Compression Ratio(in Percentage) | 38% |

| User Parameter (No. Of Bits) | 6 |
|---|---|
| Input File Size(in Bytes) | 34618 |
| Output File Size(in Bytes) | 25985 |
| Compression Ratio(in Percentage) | 25% |

| User Parameter (No. Of Bits) | 7 |
|---|---|
| Input File Size(in Bytes) | 34618 |
| Output File Size(in Bytes) | 30304 |
| Compression Ratio(in Percentage) | 13% |

| User Parameter (No. Of Bits) | 8 |
|---|---|
| Input File Size(in Bytes) | 34618 |
| Output File Size(in Bytes) | 34618 |
| Compression Ratio(in Percentage) | 0% |

### 5.2 Silence Compression Method:

1) INPUT AUDIO FILE :

Name of File:      J1. WAV
Media Length:   4.320 SEC
Audio Format:   PCM, 8000HZ, 8-BIT, Mono
File Size:             33.8KB (34,618 BYTES)

| Input File Size(in Bytes) | 34618 |
|---|---|
| Output File Size(in Bytes) | 25099 |
| Compression Ratio(in Percentage) | 28% |

2) INPUT AUDIO FILE:

Name Of File::    Chimes2.wav
Media Length:   0.63 sec
Audio Format:   PCM, 22,050Hz, 8-Bit, Mono
File Size:             14028 bytes

| Input File Size(in Bytes) | 14028 |
|---|---|
| Output File Size(in Bytes) | 7052 |
| Compression Ratio(in Percentage) | 50% |

3) INPUT AUDIO FILE:

Name Of File:    Chord2.wav
Media Length:   1.09 sec
Audio Format:   PCM, 22,050Hz, 8-Bit, Mono
File Size:             14028 bytes

| Input File Size(in Bytes) | 14028 |
|---|---|
| Output File Size(in Bytes) | 9074 |
| Compression Ratio(in Percentage) | 36% |

4) INPUT AUDIO FILE:

Name Of File:    Ding2.wav
Media Length:   0.91 sec
Audio Format:   PCM, 22,050Hz, 8-Bit, Mono
File Size:             20298 bytes

| Input File Size(in Bytes) | 20298 |
|---|---|

| Output File Size(in Bytes) | 13887 |
|---|---|
| Compression Ratio(in Percentage) | 32% |

5) INPUT AUDIO FILE:

Name Of File:   Logoff2.wav
Media Length:   3.54 sec
Audio Format :  PCM,22,050Hz,8-Bit,Mono
File Size:      783625 bytes

| Input File Size(in Bytes) | 783625 |
|---|---|
| Output File Size(in Bytes) | 60645 |
| Compression Ratio(in Percentage) | 23% |

6) INPUT AUDIO FILE:

Name Of File::  Notify2.wav
Media Length:   1.35 sec
Audio Format:   PCM, 22,050Hz, 8-Bit, Mono
File Size:      29930 Bytes

| Input File Size(in Bytes) | 29930 |
|---|---|
| Output File Size(in Bytes) | 13310 |
| Compression Ratio(in Percentage) | 56% |

## 6. Conclusions

We can achieve more compression using silence compression if more silence values are present out input audio file. Silence method is lossy method because here we consider +/-4 from 80H is consider as silence & when we perform compression we replace it with 80H runs times. When we decompress the audio file using silence method we get the original size but we do not get original data. Some losses occur. Companding method is also lossy method but one advantage is here we can adjust compression ratio according our requirement. Here depending upon no. of bits used in output file we get compression ratio. When we decompress the audio file using companding method we get the original size as well as we get original data only minor losses occur. But we get audio quality. Compare to silence method companding method is good and better.

**Kruti J Dangarwala** had passed B.E (computer science) in 2001, M.E (computer engg.) In 2005. She is currently employed at SVMIT, Bharuch, Gujarat State, India as an assistant professor. She has published two technical papers in various conferences.

**Jigar H. Shah** had passed B.E. (Electronics) in 1997, M.E. (Microprocessor) in 2006. Presently he is pursuing Ph.D. degree. He is currently employed at SVMIT, Bharuch, Gujarat State, India as an assistant professor. He has published five technical papers in various conferences and also has five book titles. He is life member of ISTE and IETE.